# Soliton solution of the generalized fifth order Hirota-Satsuma equation coupled with KdV

Gh. Forozani[1] and B.Sohrabi[2]

Department of Physics, Payame Noor University, P.O. Box 19395-3697, Tehran, Iran
e-mail: forozani@pnu.ac.ir

[2]Department of Physics, Faculty of Science, Bu-Ali Sina University, Hamedan, Iran,

**Abstract**

We introduced a fifth-order partial differential equation as a generalization of Hirota-Satsuma coupled with KdV system. This equation is investigated based on tanh method. By applying the suitable independent variable in Hirota-Satsuma equation, we can convert this partial differential equations(PDE) into ordinary differential equations(ODE) .Solving the converted Hirota-Satsuma equation by numerical methods we showed this equation have had soliton solution.

## 1. Introduction

The solitary wave and soliton phenomenon was first described in 1834 by John Scott Russell (1808–1882) when he followed the path of a solitary wave in the Union Canal in Scotland[1]. Soliton is not defined in a unique way.
Solutions of nonlinear wave equations which have the following three properties are called soliton.
1- It is confined in a finite region of space.
2- Their shape and velocity are not changed.

3- After the collision with other solitons, its shape is preserved.

Answers that include the second property, have been called Solitary waves.

The inherent stability of solitons, enable them to be sent over long distances without the use of repeaters and could potentially double transmission capacity. Soliton waves are quite stable , and in case of disturbance continue to move to its initial state.

After Russell, more than a century, the solitons were not heeded. Then in 1965, Norman Zabvsky, from Bell lab and Martin Kruskal of Princeton university, who described the behavior of solitons in terms in mathematical expression. Since then, gradually, solitons were used not only to describe water waves, but also in other fields of physics that deal with the wave and showed excellent performance.[2,3].

Nonlinear partial differential equations(PDE) in different scientific fields such as fluid mechanics, solid state physics, plasma physics, optical fibers, chemical physics [4,5] , and so on have the most importance subjects for study. Finding exact responses to these equations will help us to better understanding of our environmental nonlinear physical phenomena. For most non-linear partial differential equations, soliton solutions can be defined. One of these equations is nonlinear equation of generalized Hirota - Satsuma coupled with a kdv system which will be shown below:

$$\begin{cases} u_t = \dfrac{1}{4} u_{xxx} + 3uu_x + 3(-v^2 + w)_x & (1) \\ v_t = -\dfrac{1}{2} v_{xxx} - 3uv_x & (2) \\ w_t = -\dfrac{1}{2} w_{xxx} - 3uw_x & (3) \end{cases}$$

If in eq. **(1)** , $w$ does not depend on $x$ and $t$ , then**:** $w_x = 0$ , $w_t = 0$

The above mentioned equations can be transformed into the following equations:

$$\begin{cases} u_t = \dfrac{1}{4} u_{xxx} + 3uu_x - 6vv_x & (4) \\ v_t = -\dfrac{1}{2} v_{xxx} - 3uv_x & (5) \end{cases}$$

In this article, we introduce the generalized fifth-order of this equation which is defined as follows:

$$\begin{cases} u_t = u_{xxxxx} + \dfrac{1}{4} u_{xxx} + 3uu_x - 6vv_x & (6) \\ v_t = v_{xxxxx} - \dfrac{1}{2} v_{xxx} - 3uv_x & (7) \end{cases}$$

$u(x,t)$ and $v(x,t)$ are solutions of the above equations in (1+1) dimensions. $u_x(x,t)$ and $u_t(x,t)$ are defined as follows:

$$u_t(x,t) = \frac{\partial u(x,t)}{\partial t} \ ; \ u_x(x,t) = \frac{\partial u(x,t)}{\partial x}$$

Of course, there are different methods for finding soliton solutions of equations **(6)** and **(7)**. Some of these methods include**:**

1. Backlund transformation .[6]
2. Darboux transformation .[7]
3. Tanh-function.[8]
4. Extended tanh-function.[9]
5. Sine-cosine.[10]
6. Li group analysis and so on[11]. In this article the tanh-function method is used.

## 2. Tanh method

In this section, we find analytical solutions to equations **(6)** and **(7)** using the hyperbolic tangent method. This method is very useful for finding soliton solutions because many soliton solutions can be written as hyperbolic tangent functions. It is necessary to remember that many non-linear equations using hyperbolic tangent method has been solved and studied by A.M. Wazwaz.

If we use the independent variable $z = x - ct$ and apply it in equations **(6)** and **(7)** we have the following relationships:

$$\begin{cases} \dfrac{d^5 U(z)}{dz^5} + \dfrac{1}{4}\dfrac{d^3 U(z)}{dz^3} + 3U(z)\dfrac{dU(z)}{dz} + c\dfrac{dU(z)}{dz} - 6V(z)\dfrac{dV(z)}{dz} = 0 \quad (8) \\ \\ \dfrac{d^5 V(z)}{dz^5} - \dfrac{1}{2}\dfrac{d^3 V(z)}{dz^3} - 3U(z)\dfrac{dV(z)}{dz} + c\dfrac{dV(z)}{dz} = 0 \quad (9) \end{cases}$$

where $u(x,t) = U(z)$, $v(x,t) = V(z)$ [12].

Again we introduce another independent variable in the form $y = Tanhz$ which will lead to derivative changes as follows[13]:

$$\frac{d}{dz} = (1-y^2)\frac{d}{dy} \quad (10)$$

$$\frac{d^2}{dz^2} = (1-y^2)[-2y\frac{d}{dy} + (1-y^2)\frac{d^2}{dy^2}] \quad (11)$$

$$\frac{d^3}{dz^3} = (1-y^2)[(6y^2-2)\frac{d}{dy} - 6y(1-y^2)\frac{d^2}{dy^2} + (1-y^2)^2\frac{d^3}{dy^3}] \quad (12)$$

$$\frac{d^4}{dz^4} = (1-y^2)[-8y(3y^2-2)\frac{d}{dy} + 4(1-y^2)(9y^2-2)\frac{d^2}{dy^2} - 12y(1-y^2)^2\frac{d^3}{dy^3} + (1-y^2)^3\frac{d^4}{dy^4}] \quad (13)$$

$$\frac{d^5}{dz^5} = (1-y^2)[8(15y^4-15y^2+2)\frac{d}{dy} - 120y(1-y^2)(2y^2-1)\frac{d^2}{dy^2} + 120y(1-y^2)^3\frac{d^3}{dy^3} - 120y(1-y^2)^3\frac{d^4}{dy^4} + (1-y^2)^4\frac{d^5}{dy^5}] \quad (14)$$

With substituting **(10)** to **(14)**, in equations **(8)** and **(9)**, the result is:

$$4(1-y^2)^4 \frac{d^5U(y)}{dy^5} - 480y(1-y^2)^3 \frac{d^4U(y)}{dy^4} + (1-y^2)^2(480y^2 - 479)\frac{d^3U(y)}{dy^3} +$$

$$(1-y^2)(960y^3 + 474y)\frac{d^2U(y)}{dy^2} + (480y^4 - 474y^2 + 62 + 12U(y) + 4c)\frac{dU(y)}{dy} -$$

$$24V(y)\frac{dV(y)}{dy} = 0$$

(15)

$$2(1-y^2)^4 \frac{d^5V(y)}{dy^5} - 240y(1-y^2)^3 \frac{d^4V(y)}{dy^4} + (1-y^2)^2(240y^2 - 241)\frac{d^3V(y)}{dy^3} -$$

$$2y(1-y^2)(240y^2 - 123)\frac{d^2V(y)}{dy^2} + (240y^4 - 246y^2 + 34 - 6U(y) + 2c)\frac{dV(y)}{dy} = 0$$

(16)

For a detailed response, limited extension of $y$ is considered as following

$$U(y) = \sum_{m=0}^{M} a_m y^m \quad , \quad V(y) = \sum_{n=0}^{N} b_n y^n \tag{17}$$

Placement of **(17)** and **(10)** to **(14)** in Equations **(15)** and **(16)** and embed the highest level of linear order with the highest level of non-linear order, is determined that:

$$M = N = 4 \tag{18}$$

with placement of these numbers in the series of equation **(17)** the following equations are obtained:

$$U(y) = a_0 + a_1 y + a_2 y^2 + a_3 y^3 + a_4 y^4 \tag{19}$$

$$V(y) = b_0 + b_1 y + b_2 y^2 + b_3 y^3 + b_4 y^4 \tag{20}$$

Then, equations **(19)** and **(20)** and their derivatives in equations **(15)** and **(16)** are inserted, if the coefficients of each power of $y$ are added and equaled to zero, the following equations will be obtained.

$$a_4^2 + 280a_4 - 2b_4^2 = 0 \tag{21}$$

$$-120a_3 + 80a_2 + 7a_4 a_3 - 14b_4 b_3 = 0 \tag{22}$$

$$-5430a_4 - 160a_2 + 3a_3^2 - 6b_3^2 + 6a_2 a_4 - 12b_2 b_4 = 0 \tag{23}$$

$$-119a_3 + 8a_1 + a_1 a_4 + a_2 a_3 - 2b_1 b_4 - 2b_2 b_3 = 0 \tag{24}$$

$$9376a_4 + 3a_2 + 6a_1 a_3 + 3a_2^2 + 6a_0 a_4 - 6b_2^2 - 12b_0 b_4 - 12b_1 b_3 + 2a_4 c = 0 \tag{25}$$

$$1943a_3 - 79a_1 + 6a_2 a_1 + 6a_0 a_3 - 12b_0 b_3 - 12b_1 b_2 + 2a_3 c = 0 \tag{26}$$

$$5754a_4 + 268a_2 + 3a_1^2 + 6a_0 a_2 - 12b_0 b_2 - 6b_1^2 = 0 \tag{27}$$

$$-1437a_3 + 31a_1 + 6a_1 a_0 + 2a_1 c + 4a_2 c - 12b_0 b_1 = 0 \tag{28}$$

$$760b_4 - a_4 b_4 = 0 \tag{29}$$

$$840b_3 - 3a_4 b_3 - 4a_3 b_4 = 0 \tag{30}$$

$$-7380b_4 + 240b_2 - 2a_4 b_2 - 4a_2 b_4 - 3a_3 b_3 = 0 \tag{31}$$

$$-1570b_3 + 40b_1 - a_4 b_1 - 2a_3 b_2 - 3a_2 b_3 - 4a_1 b_4 = 0 \tag{32}$$

$$18848b_4 - 726b_2 - 246b_3 - 3a_3 b_1 - 6a_2 b_2 + 4b_4 c - 9a_1 b_3 - 12a_0 b_4 = 0 \tag{33}$$

$$985b_3 - 41b_1 - a_2 b_1 - 2a_1 b_2 - 3a_0 b_3 + b_3 c = 0 \tag{34}$$

$$-5772b_4 + 280b_2 - 3a_1 b_1 - 6a_0 b_2 = 0 \tag{35}$$

$$-723b_3 + 17b_1 - 3a_0 b_1 + b_1 c + 2b_1 c = 0 \tag{36}$$

We can solve equations **(21)** to **(36)** and the coefficients are obtained as follows:

**Case 1:**

$$\begin{cases} a_0 = 1.3783 \times 10^{12} \\ a_1 = -3.67716 \times 10^{10} \\ a_2 = -32988036 \\ a_3 = 226205.01 \\ a_4 = 760 \end{cases} , \quad \begin{cases} b_0 = 1.66474 \times 10^{12} \\ b_1 = 1012811 \times 10^{10} \\ b_2 = 150075 \times 10^3 \\ b_3 = -166205 \\ b_4 = 628.65 \end{cases} , \quad c = 4.02774 \times 10^{12} \tag{37}$$

By substitutions these coefficients in equations **(19)** and **(20)** the soliton solutions can be obtained as follows, of course $y = Tanh z$ and $z = x - ct$:

$$u_1(x,t) = 1.3783 \times 10^{12} - 3.67716 \times 10^{10} \tanh(x - 4.02774 \times 10^{12} t) - 32988036 \tanh^2 (x - 4.2774 \times 10^{12} t) + 22620501 \tanh^3 (x - 4.02774 \times 10^{12} t) + 760 \tanh^4 (x - 4.02774 \times 10^{12} t)$$

$$\tag{38}$$

$$v_1(x,t) = 1.66474 \times 10^{12} - 1.12811 \times 10^{10} \tanh(x - 4.02774 \times 10^{12} t) - 150075000 \tanh^2$$
$$(x - 4.2774 \times 10^{12} t) - 166205 \tanh^3(x - 4.02774 \times 10^{12} t) + 628.65 \tanh^4(x - 4.02774 \times 10^{12} t)$$

**(39)**

The numerical simulation of two soliton solutions $u_1$ and $v_1$, are presented in the following figures 1, 2, 3 and 4

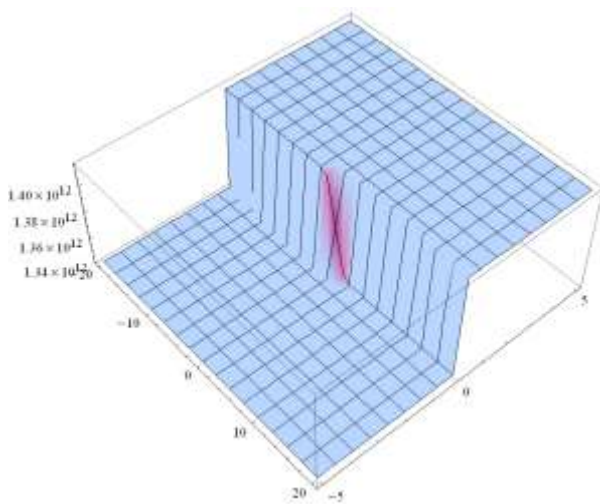

Fig.1.u₁(x,t)

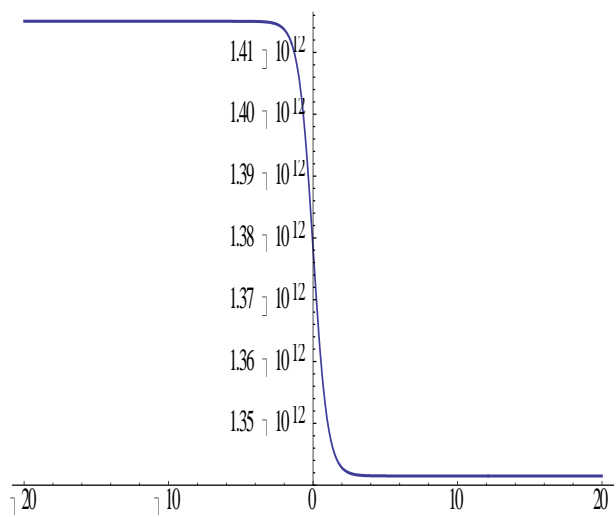

Fig.2.u₁(x,t) when t=0

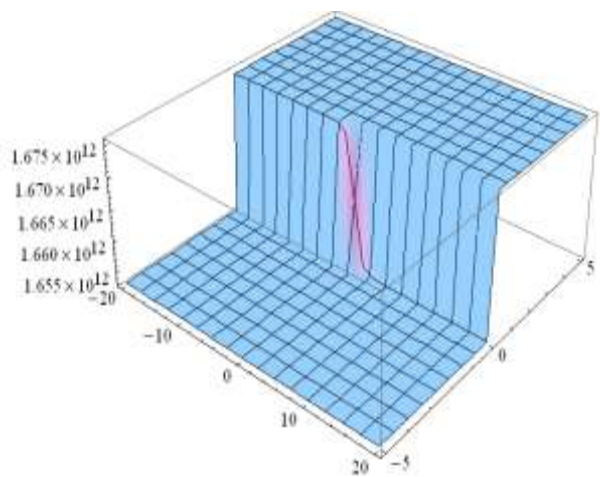

Fig.3.v₁(x,t)

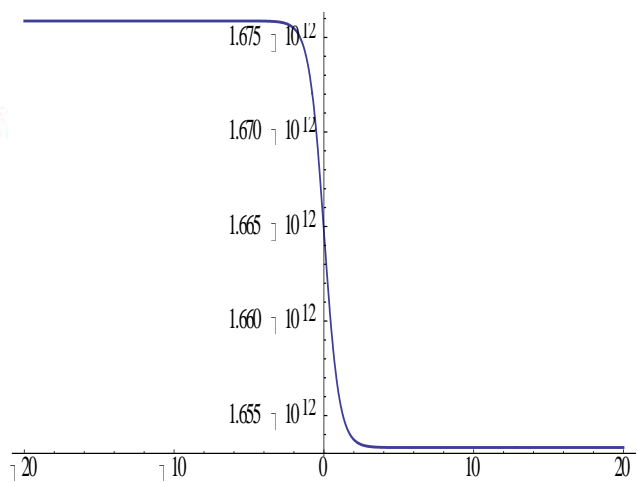

Fig.4.v₁(x,t) when t=0

**Case 2:**

$$\begin{cases} a_0 = -879.661 \\ a_1 = 30574.3 \\ a_2 = -14091.938 \\ a_3 = 96.631 \\ a_4 = 760 \end{cases} , \quad \begin{cases} b_0 = 454.472 \\ b_1 = -20440.2 \\ b_2 = -336671 \\ b_3 = -71 \\ b_4 = 628.65 \end{cases} , \quad c = -617.0864 \quad (40)$$

By using these coefficients in equations **(19)** and **(20)** soliton solutions can be obtained as follows

$$u_2(x,t) = -879.661 + 30574.3\tanh(x+617.0864t) - 14091.938\tanh^2(x+617.0864t) - 96.631\tanh^3(x+617.0864t) + 760\tanh^4(x+617.0864t) \quad (41)$$

$$v_2(x,t) = 454.472 - 20440.2\tanh(x+617.0864t) - 336671\tanh^2(x+617.0864t) - 71\tanh^3(x+617.0864t) + 628.65\tanh^4(x+617.0864t) \quad (42)$$

The numerical simulation of two solitons $u_2$ and $v_2$, are presented in the figures 5, 6, 7 and 8.

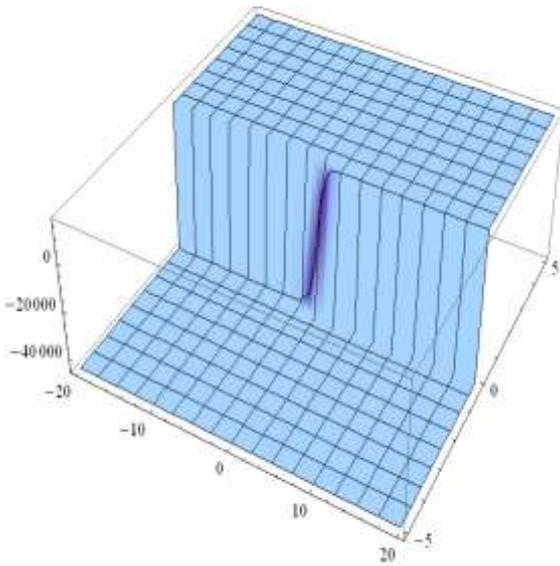

Fig. 5 $u_2(x,t)$

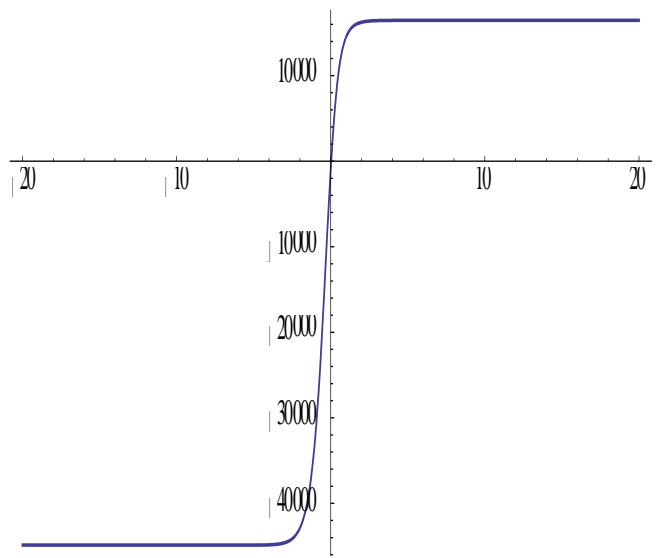

Fig. 6 $u_2(x,t)$ when t=0

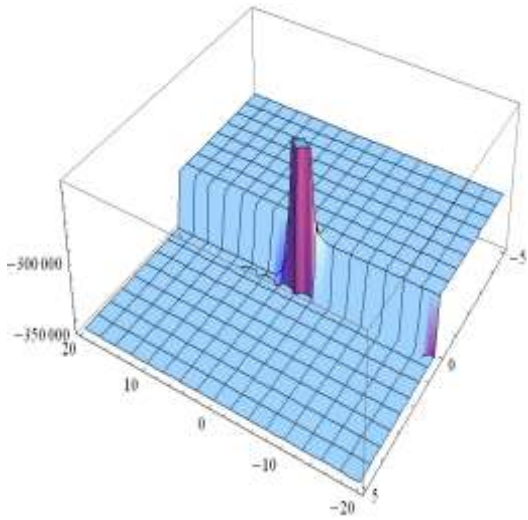
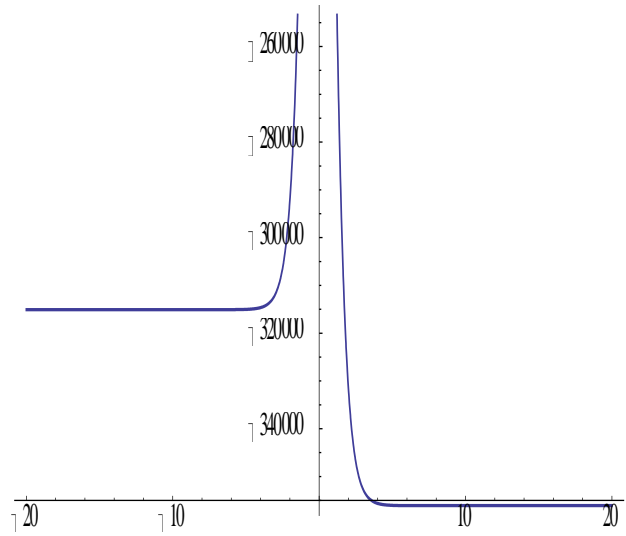

Fig. 7 v₂(x,t)         Fig. 8 v₂(x,t) when t=0

If it is assumed that $b_3 = 0$ we have:

$$\begin{cases} a_0 = 3619.13 \\ a_1 = 0 \\ a_2 = 0 \\ a_3 = 0 \\ a_4 = 760 \end{cases} \quad \begin{cases} b_0 = -666.21 \\ b_1 = 0 \\ b_2 = -547 \\ b_4 = 628.65 \end{cases} \quad c = 5987.46 \tag{43}$$

Making these substitutions in equations **(18)** and **(19),** will give the soliton solutions as follows:

$$u_3(x,t) = 3619.3 + 760 \tanh^4(x - 5987.46t) \tag{44}$$

$$v_3(x,t) = -666.21 - 547 \tanh^2(x - 5987.46t) + 628.65 \tanh^4(x - 5987.46t) \tag{45}$$

The numerical simulation of two solitons $u_3$ and $v_3$, are presented in figures 9, 10, 11 and 12.

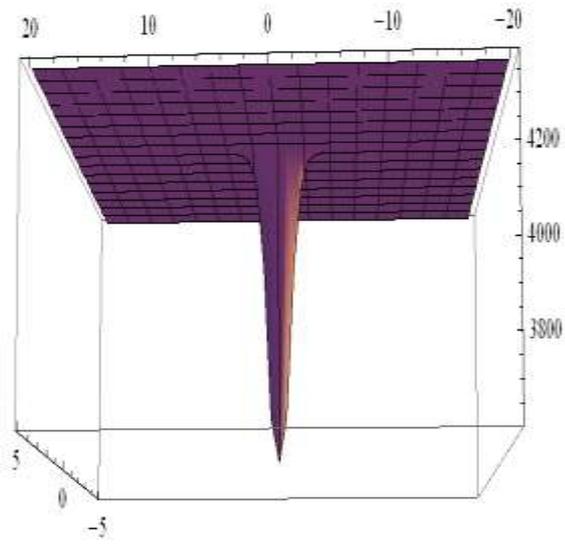 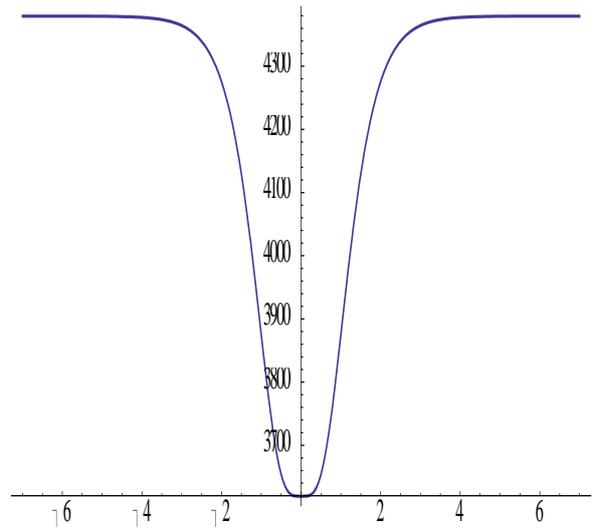

Fig. 9 $u_3(x,t)$  Fig.10 $u_3(x,t)$ when $t=0$

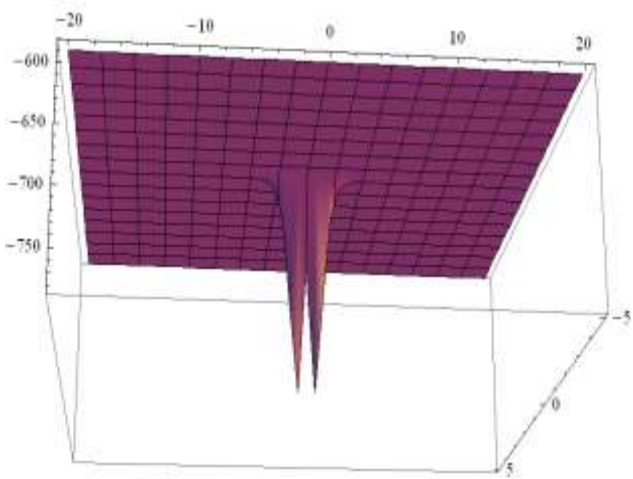 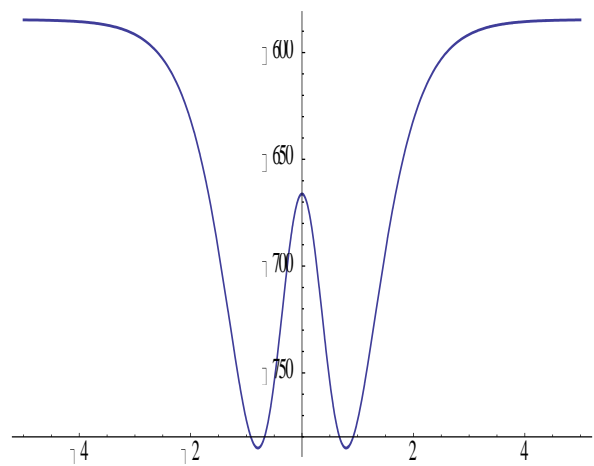

Fig. 11 $v_3(x,t)$  Fig. 12 $v_3(x,t)$ when $t=0$

## 3. Conclusions

The main goal in this article is find the soliton solutions of fifth-order nonlinear equation of the generalized Hirota-Satsuma coupled with the kdv system (1+1) dimensional and since most forms of soliton equations are hyperbolic tangent functions we used hyperbolic tangent method.

## 4. Acknowledgements

Author was supported by Department of Physics, Payame Noor University, Shiraz Iran.